\documentclass[epj]{svjour}
\usepackage{graphics}
\begin{document}
\title{The $^{27}$Al(p,$\alpha$)$^{24}$Mg reaction at astrophysical energies studied by means of the Trojan Horse Method applied to the $^2$H($^{27}$Al,$\alpha^{24}$Mg)n reaction}
\author{S. Palmerini\inst{1,2} \and 
M. La Cognata\inst{3} \and 
F. Hammache\inst{4} \and
L. Acosta\inst{5} \and
R. Alba\inst{3} \and
V. Burjan\inst{6} \and 
E. Ch\'avez\inst{5} \and
S. Cherubini\inst{7,3} \and
A. Cvetinovi\'c\inst{8} \and
G. D'Agata\inst{6} \and
N. de S\'er\'eville\inst{4} \and
A. Di Pietro\inst{3} \and
P. Figuera\inst{3} \and
Z. F\"ul\"op\inst{10} \and
K. Gait\'an De Los Rios\inst{5} \and
G.L. Guardo\inst{3,7} \and
M. Gulino\inst{11,3} \and
S. Hayakawa\inst{12} \and
G.G. Kiss\inst{10} \and
M. La Commara\inst{13,14} \and
L. Lamia\inst{7,3} \and
C. Maiolino\inst{3} \and
G. Manic\'o\inst{7,3} \and
C. Matei\inst{15} \and
M. Mazzocco\inst{16,17} \and
J. Mrazek\inst{6} \and
T. Parascandolo\inst{14} \and
T. Petruse\inst{15,18} \and
D. Pierroutsakou\inst{14} \and
R.G. Pizzone\inst{3} \and
G.G. Rapisarda\inst{7,3} \and
S. Romano\inst{6,3} \and
D. Santonocito\inst{3} \and
M.L Sergi\inst{7,3} \and 
R. Spart\`a\inst{3,7} \and
A. Tumino\inst{11,3} \and 
H. Yamaguchi\inst{12} 
%
%
%
}                     
\offprints{}          
\institute{Dipartimento di Fisica e Geologia, Universit\`a  degli Studi di Perugia, 06123 Perugia, Italy \and 
Istituto Nazionale di Fisica Nucleare, Sezione di Perugia, 06123 Perugia, Italy \and 
Istituto Nazionale di Fisica Nucleare, Laboratori Nazionali del Sud, 95123 Catania, Italy \and 
Institut de Physique Nucl\'eaire d'Orsay, UMR8608, IN2P3-CNRS, Universit\'e Paris Sud 11, 91406 Orsay, France \and 
Instituto de F\'isica, Universidad Nacional Aut\'onoma de Mexico, A.P. 20-364, 01000 Mexico City, Mexico\and
Nuclear Physics Institute of the Czech Academy of Sciences, 250 68 $\check{{\rm R}}$e$\check{{\rm z}}$, Czech Republic \and 
Dipartimento di Fisica e Astronomia ``Ettore Majorana'', Universit\`a degli Studi di Catania, Catania 95123, Italy \and 
Jo$\check{{\rm z}}$ef Stefan Institute, Jamova cesta 39, Ljubljana, Slovenia \and
Institute for Nuclear Research (ATOMKI), H-4001 Debrecen, POB.51, Hungary \and 
Facolt\`a di Ingegneria e Architettura, Universit\`a degli Studi ``Kore", 94100 Enna, Italy \and
Center for Nuclear Study, University of Tokyo, RIKEN Campus, Wako, Saitama 351-0198, Japan \and
Dipartimento di Farmacia, Universit\`a degli Studi di Napoli ``Federico II'', 80131 Napoli, Italy \and 
Istituto Nazionale di Fisica Nucleare, Sezione di Napoli, 80126 Napoli, Italy \and 
Extreme Light Infrastructure - Nuclear Physics (ELI-NP)/Horia Hulubei National R\&D Institute for Physics and Nuclear Engineering (IFIN-HH), Bucharest-Magurele, Romania \and 
Dipartimento di Fisica e Astronomia, Universit\`a degli Studi di Padova, 35131 Padova, Italy \and 
Istituto Nazionale di Fisica Nucleare, Sezione di Padova, 35131 Padova, Italy \and 
Scoala Doctorada de Ingineria si Aplicatiile Laserilor si Acceleratorilor, Politechnica University, Bucharest, Romania
}

\date{Received: date / Revised version: date}
%
\abstract{
The $^{27}$Al(p,$\alpha$)$^{24}$Mg reaction, which drives the destruction of $^{27}$Al and the production of $^{24}$Mg in stellar hydrogen burning, has been investigated via the Trojan Horse Method (THM), { by measuring the $^2$H$(^{27}$Al,$\alpha^{24}$Mg)n three-body reaction. The experiment covered a broad energy range 
($E_{\rm c.m.}$ $\le$ 1.5 MeV), aiming to investigate those of interest for astrophysics.}  The  results confirm the THM as a valuable technique for the experimental study of fusion reactions at very low energies and { suggests} the presence of a rich pattern of resonances in the energy region close to the Gamow window of stellar hydrogen burning (70-120 keV), with potential impact on astrophysics. { To estimate such an impact a second run of the experiment is needed, since the background due the three-body reaction hampered to collect enough data to resolve the resonant structures and extract the reaction rate.}
\PACS{
      {PACS-key}{discribing text of that key}   \and
      {PACS-key}{discribing text of that key}
     } 
} 
\maketitle
\section{Introduction}
\label{intro}
Aluminum has only one stable isotope, with A = 27, but in astrophysics it is the unstable isotope with A = 26 (and half-life $7.17 \times 10^5$~years) that arouses the greatest interest in the scientific community. { The isotope is believed to provide the heat needed to differentiate the interiors of the small planetary bodies, but despite its importance in the formation of the rocky planets its stellar source is still debated.
On the one side massive objects and supernova progenitors 
are the most likely candidates for this role, but the $^{26}$Al fossil abundances  would instead indicate that the main producers of Al-rich dust are low- and  intermediate-mass stars during the asymptotic giant branch (AGB) phase \cite{kl14,ventura16,pal17}.

A widely-distributed $\gamma$ emission at 1809~keV, following the isotope decay into excited states of $^{26}$Mg, 
proves that $^{26}$Al nucleosynthesis is active today in the Milky Way. 
The first clear detection of 1.808 MeV gamma lines from the bulge of the galaxy was made by the HEAO-3 satellite in 1984 \cite{Mah84}, while the first mapping of the $^{26}$Al emission in the Milky Way was due to the COMPTEL satellite \cite{diehl}.
Nowadays, modern millimiter and submillimeter interferometer arrays,
as e.g. ALMA (Atacama Large Millimeter/submillimeter Array) and NOEMA (Northern Extended Millimeter Array), are able to spatially identify discrete objects which are active sources of  $^{26}$Al in the Galaxy.
However, only a few objects have been observed and they have not yet allowed to unambiguously identify the main galactic source of 26Al.
This is, for instance, the case of the detection of the  molecule $^{26}$AlF in the nova remnant CK Vul \cite{kami}. The observation itself is a great success of stellar spectroscopy and the estimated abundance of $^{26}$Al indicates that one of the merging objects was at least one solar-mass red giant star, but  the red nova rate is too small to think that objects like CK Vul are the major producers of galactic $^{26}$Al.

}


The fossil abundance of $^{26}$Al  is suggested by a superabundance of 
its daughter nucleus, $^{26}$Mg, in comparison with the most abundant Mg 
isotope (A = 24) in meteorites. From the ratios of the abundances 
$^{26}$Mg/$^{24}$Mg and $^{26}$Mg/$^{27}$Al measured in presolar dust, 
meteorites and early solar system materials it is possible to estimate 
the $^{26}$Al/$^{27}$Al ratio in the ancient Galaxy and to date these 
ancient solids \cite{vescovi18}.  For this reason, it is crucial to 
know with high precision the nucleosynthesis process not only of 
$^{26}$Al but also of $^{27}$Al and $^{24}$Mg as well. In particular, 
all these nuclei take part to the so called MgAl cycle, typical of 
high-temperature (T=0.055~GK or T$_9$=0.055 in units of GK) H-burning 
of evolved stars (see \cite{ili11} for an extensive discussion on the 
role of aluminum isotopes). 

Because of the higher Coulomb barriers involved, the MgAl cycle is 
not as relevant as the CNO one for energy production in stars, but 
plays an important role in the nucleosynthesis of Al and Mg. In this 
framework, the $^{27}$Al(p,$\alpha$)$^{24}$Mg reaction drives the 
destruction of $^{27}$Al, the production of $^{24}$Mg and closes the MgAl 
cycle when its rate exceeds the one of the competing $^{27}$Al(p,$\gamma$)$^{28}$Si 
reaction. However, at temperatures T$_9<$0.1 it is difficult to compare 
the $^{27}$Al(p,$\alpha$)$^{24}$Mg reaction with the competitor  
$^{27}$Al(p,$\gamma$)$^{28}$Si channel because of the uncertainties, 
which are so large to make astrophysical predictions unreliable \cite{ili11}.

The most recent published rate for the $^{27}$Al(p,$\alpha$)$^{24}$Mg reaction 
\cite{ili10} is widely employed in astrophysical calculations, but at  
T$_{9}\le$ 0.1 its lower, median and upper values are 1.85$\times$10$^{-11}$, 
4.34$\times$10$^{-11}$ and 8.51$\times$10$^{-11}$ cm$^{3}$mol$^{-1}$s$^{-1}$, 
respectively, spanning almost one order of magnitude. The uncertainty range becomes larger at lower temperatures.
{
Such reaction rate is based on direct measurements data \cite{Timmermann88} (spanning only the low-energy region, between 200 and 360~keV, and setting upper limits for some resonance strengths), on spectroscopic data \cite{Endt1990,Endt1998}, on transfer-reaction data \cite{Champ88} and on shell-model  \cite{Endt93}.}

Nowadays, the abundances of magnesium isotopes in the interstellar medium are considered a powerful probe of star formation processes over cosmological time-scales. This consideration is based on the assumption that the main contribution of $^{24}$Mg comes from massive stars whereas $^{25}$Mg and $^{26}$Mg come from intermediate mass objects \cite{van19}, but this scenario could be modified, strengthened or weakened by more accurate measurements of proton capture reactions on Al and Mg isotopes in the energy range of stellar nucleosynthesis. { For example, if the rate of the $^{27}$Al(p,$\alpha$)$^{24}$Mg reaction were found to be higher at low temperatures, the production of $^{24}$Mg could  be more efficient even in intermediate mass objects (5-8 M$_\odot$).}

{ Meteoritic grain abundances are among the most precise constrains available for nucleosynthesis studies because of their relatively small uncertainties, which in the case of magnesium and aluminum isotopic ratios are on average 10\% and 15\%, respectively.  

However, in the last years, important information about Mg and Al nucleosynthesis are coming also from} high-resolution stellar spectroscopy. It has shown that the already known  anti-correlation of Mg-Al abundances of red-giant-branch stars of globular clusters (e.g. $\omega-Cen$, M4, NGC 2808)  hides the existence of multiple stellar populations, and that, in some cases, the relative abundances of $^{24}$Mg, $^{25}$Mg and $^{26}$Mg do not show any obvious correlation with Al abundances \cite{lind,dacosta}. To account for these observations a complicate scenario with various polluters (massive fast rotating stars, intermediate mass AGB and super AGB stars) is required \cite{carretta18}, along with quite specific assumptions in the theoretical models, because in the narrow T$_9$ range between 0.07 and 0.08  the temperature of the stellar H-burning makes the difference among producing, saving or destroying $^{24}$Mg \cite{yong03}.
Therefore, it is crucial to measure with high precision the 
$^{27}$Al(p,$\alpha$)$^{24}$Mg reaction rate, as well as the 
rates of all the reactions involved in the MgAl cycle in the energy range 
typical of stellar nucleosynthesis.

Finally, it is worth noting that if resonances of the $^{27}$Al(p,$\alpha$)$^{24}$Mg 
reaction will be found at low energies, also the contribution of low-mass-star 
H-burning to the Al nucleosynthesis should be revised.


\section{Experimental method and set-up}
\label{thm}

\begin{figure}
\resizebox{0.48\textwidth}{!}{
\includegraphics{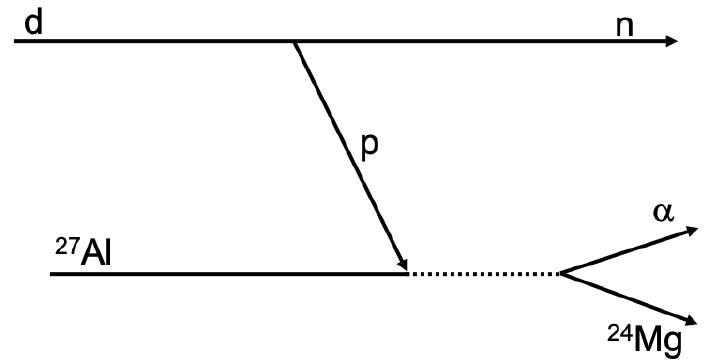}}
\caption{Sketch of the QF transfer reaction to be selected among all the data recorded during the experiment.}
\label{dia}       
\end{figure}

The $^{27}$Al(p,$\alpha$)$^{24}$Mg reaction has been investigated 
using the Trojan Horse Method (hereafter THM), which is an experimental 
indirect technique already successfully applied to study several astrophysically 
relevant reactions by using appropriate three-body quasi-free (QF) processes 
\cite{tribble14,spit19}. The method has proven to be particularly suited for acquiring 
information on charged-particle induced reaction cross sections at 
astrophysical energies, since it allows one to overcome the Coulomb 
barrier in the two-body entrance channel. In particular, many (p,$\alpha$) 
reactions involved in low- and high-temperature H-burning networks have 
been measured, including a few involving radioactive nuclei (see e.g. \cite{laco17}  and \cite{cherubini15} and references therein for the $^{18}$F$(p,\alpha)^{15}$O) and neutrons (see e.g.\cite{guardo17} for the $^{17}$O$(n,\alpha)^{14}$C, \cite{lamia19} for the $^7$Be$(n,\alpha)^4$He and \cite{pizzone20} for the $^3$He(n,p)$^3$H reaction).

One can briefly describe the method as follows. A projectile $a$ hits a
target nucleus $A$, whose wave function has a large amplitude for a $s-b$ 
cluster configuration. Under proper kinematic conditions, the particle 
$a$ interacts only with the part $b$ (participant) of the target nucleus 
$A$, while the other part $s$ behaves as a spectator to the process 
$a+b(+s) \rightarrow c+d(+s)$, which is a quasi-free (QF) mechanism. 
Since the bombarding energy is chosen to overcome the Coulomb barrier in 
the entrance channel for the $a+A \rightarrow c+d+s$ reaction, the particle 
$b$ can be brought into the nuclear field to induce the $a+b \rightarrow c+d$ 
reaction. Moreover, since the beam energy can be compensated for by the 
$b+s$ binding energy, the two-body reaction can take place at very low
$a-b$ relative energy, e.g. in the region of astrophysical relevance. 

{ We studied the $^{27}$Al(p,$\alpha$)$^{24}$Mg reaction 
($Q_2=1.6$~MeV) via the  $^2$H$(^{27}$Al,$\alpha^{24}$Mg)n three-body reaction ($Q_3=-0.6$~MeV). As it is shown in Fig. \ref{dia}, the $^{27}$Al is the projectile nucleus $a$ impinging onto a deuteron (the target nucleus A), which acts as THM nucleus because of its obvious $p-n$ structure, being the neutron the spectator nucleus $s$.

To perform an experiment aimed to investigate the  
$^2$H$(^{27}$Al,$\alpha^{24}$Mg)n three-body reaction in the $^{27}$Al$-p$ energy range between the threshold and $\sim$ 1.5~MeV, thus covering the region of astrophysical importance and a broad energy region  where direct measurements are available for normalization, we used a 60~MeV $^{27}$Al beam, delivered by the INFN-LNS Tandem (Catania, Italy), impinging onto a CD$_2$ target (isotopically enriched to 98$\%$) 120~${\rm\mu}$g$/$cm$^2$ thick, placed at 90${^\circ}$ with respect to the beam axis. 
In this framework, the deuteron is an ultimate THM nucleus thanks to its $p-n$ structure, its low‐binding energy ($B_{p-n}\sim$2.2 MeV), and a 
very well-known $l=0$ $p-n$ momentum distribution given in terms of the radial Hulth\'en wave function (see \cite{lamia12} for more details on the $p-n$ wave function).

In the plane wave impulse approximation, the three-body cross section can be expressed as: 
\begin{equation}\label{thmeq}
\frac{d^3\sigma}{dE_{\alpha}d\Omega_{\alpha}d\Omega_{^{24}Mg}}\propto(KF)\mid\Phi(p_n)\mid^2\left(\frac{d\sigma}{d\Omega}\right)^{HOES}
\end{equation}
where $KF$ is a kinematical factor containing the final state phase space 
factor, $|\Phi(p_n)|^2$ is the momentum distribution of the spectator neutron 
inside the deuteron, and $\left(\frac{d\sigma}{d\Omega}\right)^{HOES}$ is the differential half-off-energy-shell (HOES) cross section for the two-body reaction $^{27}$Al(p,$\alpha$)$^{24}$Mg, the transferred proton being virtual (see, e.g. 
\cite{spitaleri2019} and references therein for more details).
The deduced two-body reaction cross section $\left(\frac{d\sigma}{d\Omega}\right)^{HOES}$ essentially 
represents the nuclear part alone, the Coulomb barrier being already 
overcome in the entrance channel, devoid of electron screening effects.}

In the case of resonant reactions, a more advanced approach has been
developed, to account for the HOES nature of the THM cross section
and for the effect of energy resolution, both in the case of narrow
\cite{laco10,tum18} and broad resonances \cite{laco10b}, firstly introduced
by A.M. Mukhamedzhanov. Since the inspection of the resonance list
in \cite{ili10} shows that the low-energy cross section of the 
$^{27}$Al(p,$\alpha$)$^{24}$Mg is dominated by narrow resonances, in 
comparison with the typical THM energy resolution of few ten of keV (FWHM),
the application of the formalism thoroughly discussed in \cite{laco10}
will lead to the determination of the resonance strengths from the 
THM cross section. A drawback of the method is the need to normalize 
the deduced resonance strengths to a reference one reported in the literature.
However, it has been shown that extending the normalization procedure to
more than one resonance greatly reduces possible systematic errors 
(see \cite{laco15} for details).


\begin{figure}
\resizebox{0.48\textwidth}{!}{
\includegraphics{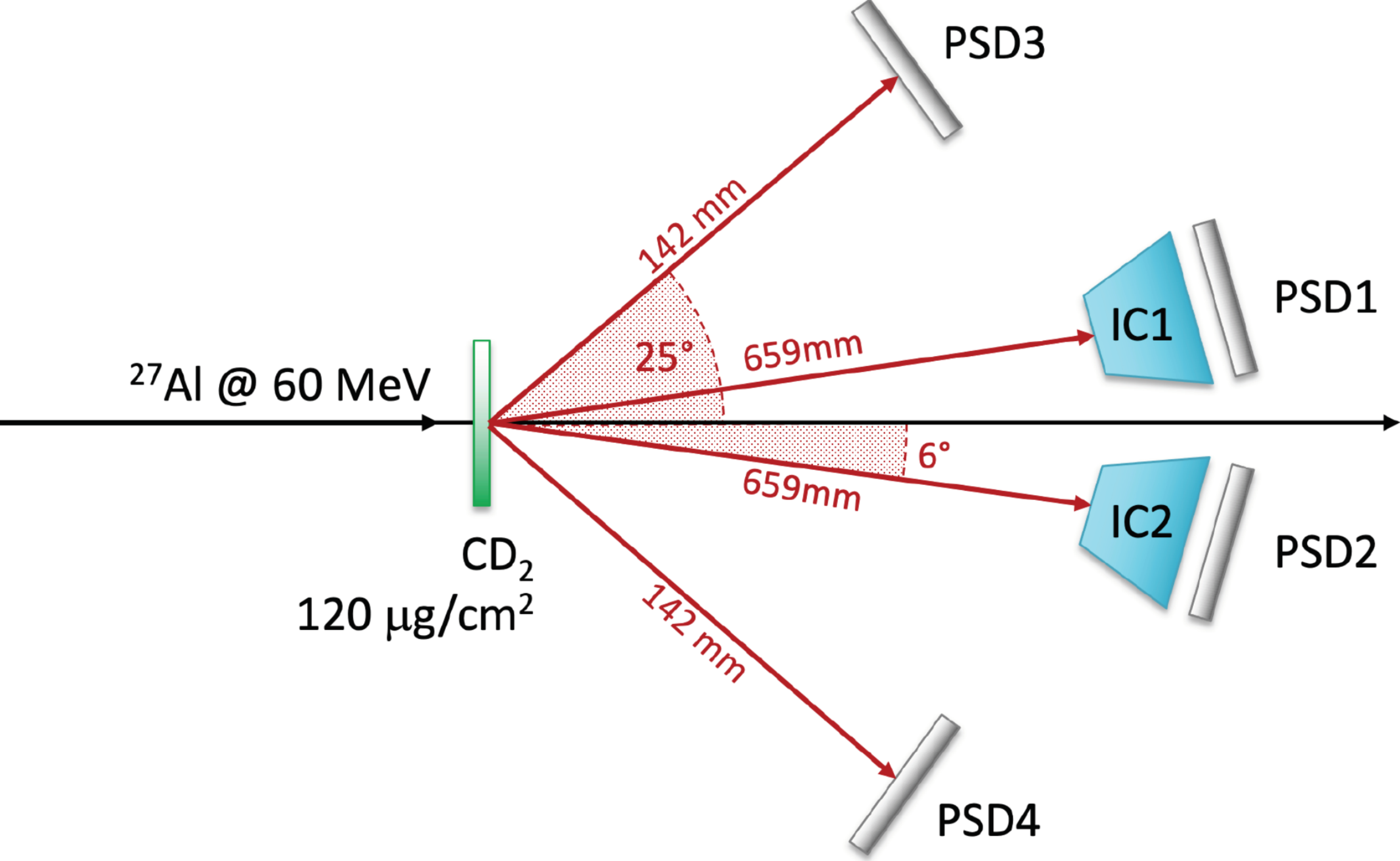}}
\caption{Sketch of the experimental set-up.}
\label{setup}       
\end{figure}

In the case of emission of three particles in the reaction, only two of them have to 
be detected and identified. Indeed, measuring their energies and emission angles, the kinematic properties of the reaction are completely determined, so there is
no need to use neutron detectors to observe the candidate spectator neutron.
The experimental set-up was optimized to detect the ejected $\alpha$ particle, 
that is, the lighter charged fragment, in coincidence with the  $^{24}$Mg recoil. Moreover, the detector positions were chosen to span the candidate $^2{\rm H}(^{27}$Al,$\alpha^{24}$Mg)n QF kinematic region, whose occurrence is a necessary condition for the THM applicability.

\begin{figure}
\resizebox{0.48\textwidth}{!}{
\includegraphics{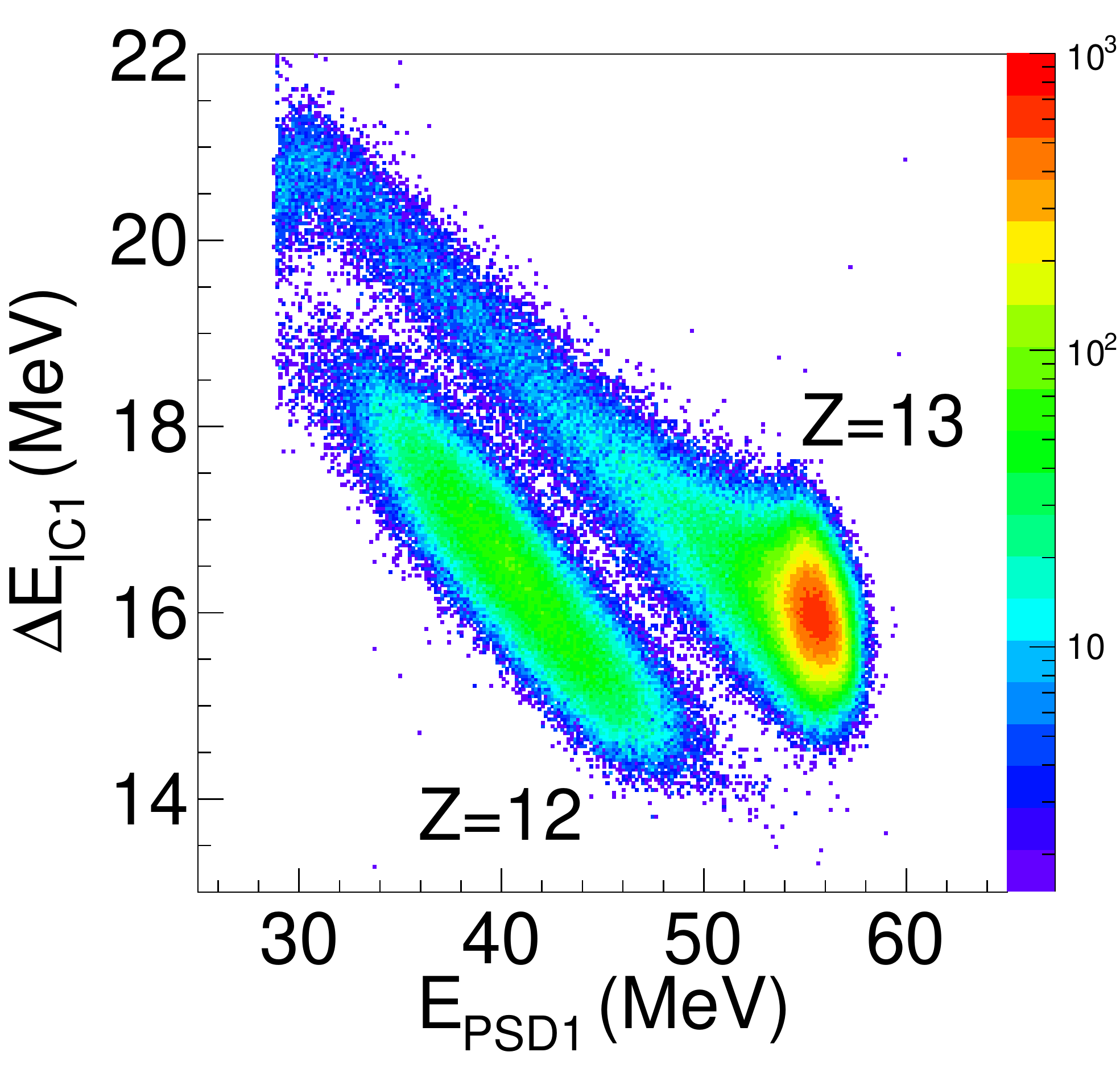}}
\caption{2D $\Delta$E-E spectrum for the telescope made up of the IC1 and the PSD1, detecting the energy loss and the residual energy
of the impinging ions, respectively. The loci for Mg (Z=12) and Al (Z=13) isotopes can be easily identified thanks to their neat separation.}
\label{de}       
\end{figure}

Fig. \ref{setup} shows a sketch of the experimental set-up we adopted. It was 
symmetric with respect to the beam axis and consisted of two telescopes and two 
silicon position sensitive detectors (PSD). Both telescopes covered an angular 
range from 4${^\circ}$ to 8${^\circ}$ on the left and on the right side of 
the beam axis.  They were made up of an ionization chamber (IC) and a PSD 
to identify Z=12 ions, in particular to discriminate between Mg and Al nuclei 
via the $\Delta$E-E technique. The IC were filled with 65 mbar isobutane gas that provided an energy resolution of about 10$\%$. A picture of the calibrated
$\Delta$E-E 2D spectrum for the IC1-PSD1 telescope is shown in Fig.\ref{de},
where the atomic number of the identified ions is marked. The red spot in the
figure is linked to the $^{27}$Al ions scattered off from carbon in the CD$_2$ targets.
The other two silicon PSDs, used to detect the $\alpha$ particles in coincidence { with $^{24}$Mg}, covered about 20${^\circ}$ from 15${^\circ}$ to 35${^\circ}$ on 
the left and on the right of the beam line. The four 1000-micron PSDs had 
$50 \times 10~{\rm mm}^2$ sensitive area with 0.5~mm position resolution and 
an energy one of 0.5$\%$ (FWHM).

{ Energy and position calibration of the detectors spanning 15${^\circ}$ to 35${^\circ}$ was performed using the $\alpha$ particles from the $^6$Li($^{12}$C,$\alpha$)$^{14}$N reaction induced by a $^6$Li beam at 8 MeV impinging on a CH$_2$ target. A standard three-peak alpha calibration source was also used for the low energy part of the $\alpha$ spectrum. The PSDs sitting at more forward angles and the ICs were calibrated by using scattering off Au and off $^{12}$C of $^24$Mg beams at 40, 45, 50 and 55~MeV.}

\section{Data analysis}
\label{analysis}

Discrimination of Mg ions from other reaction and scattering products is the
first step in the analysis, aiming at separating the reaction channel of
interest, namely, the $^2$H$(^{27}$Al,$\alpha^{24}$Mg)n three-body reaction,
from other processes taking place in the target. From the analysis of
reaction kinematics, it is apparent that the use of ICs as $\Delta$E 
detectors did not introduce detection thresholds on the Mg energy spectra.
Conversely, $\alpha$-particles energy spectra would have been significantly affected.
\begin{figure}
\resizebox{0.48\textwidth}{!}{
\includegraphics{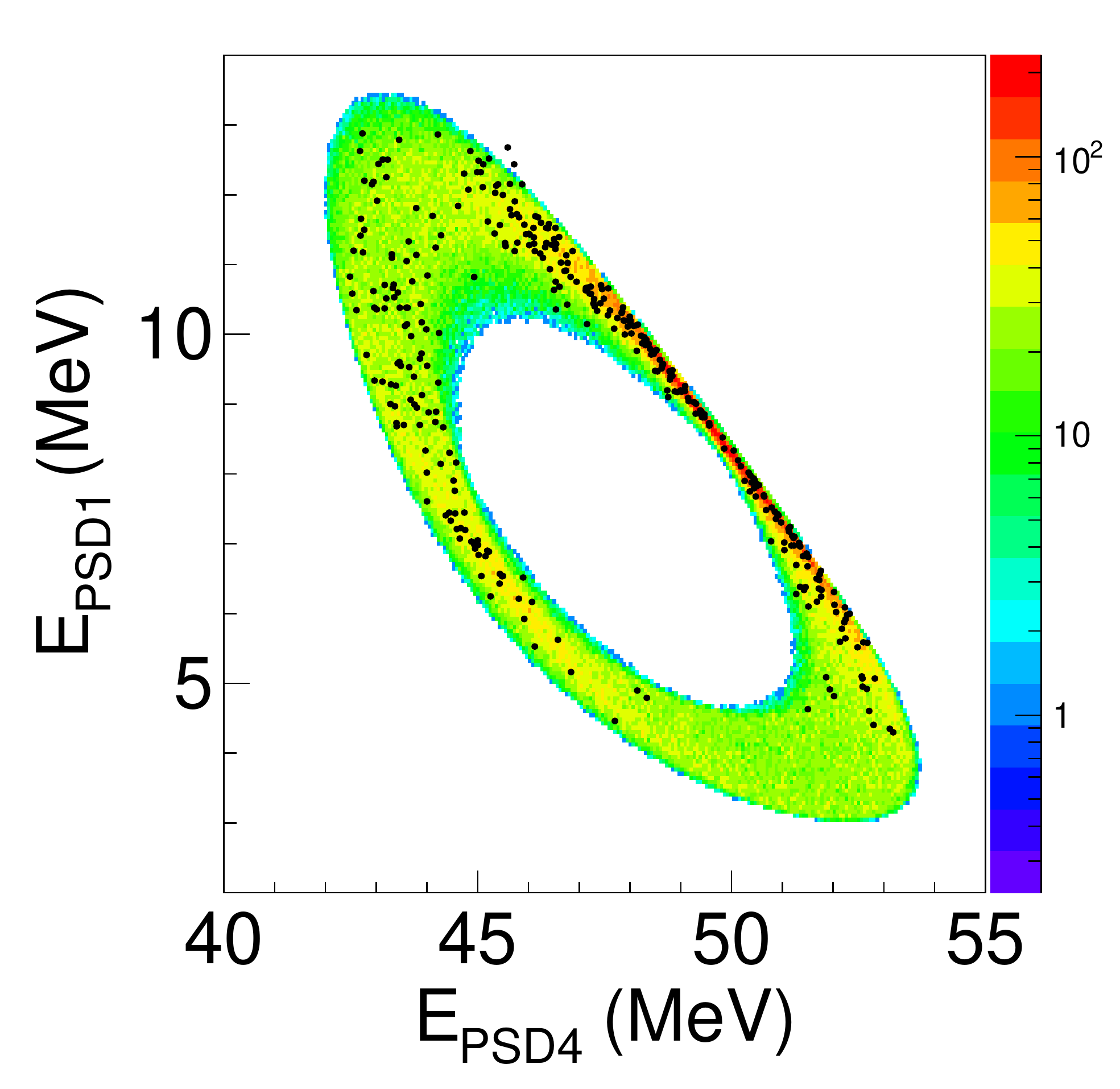}}
\caption{${\rm E}_{{\rm PSD1}}-{\rm E}_{{\rm PSD4}}$ 2D spectrum for
$4.1{^\circ}\le\theta_{{\rm PSD1}}\le5.1{^\circ}$
and $27{^\circ}\le\theta_{{\rm PSD4}}\le30{^\circ}$. THM
data are shown in black, while the Monte Carlo simulated
spectrum is shown in colors.}
\label{kinloc}       
\end{figure}
\begin{figure}
\resizebox{0.48\textwidth}{!}{
\includegraphics{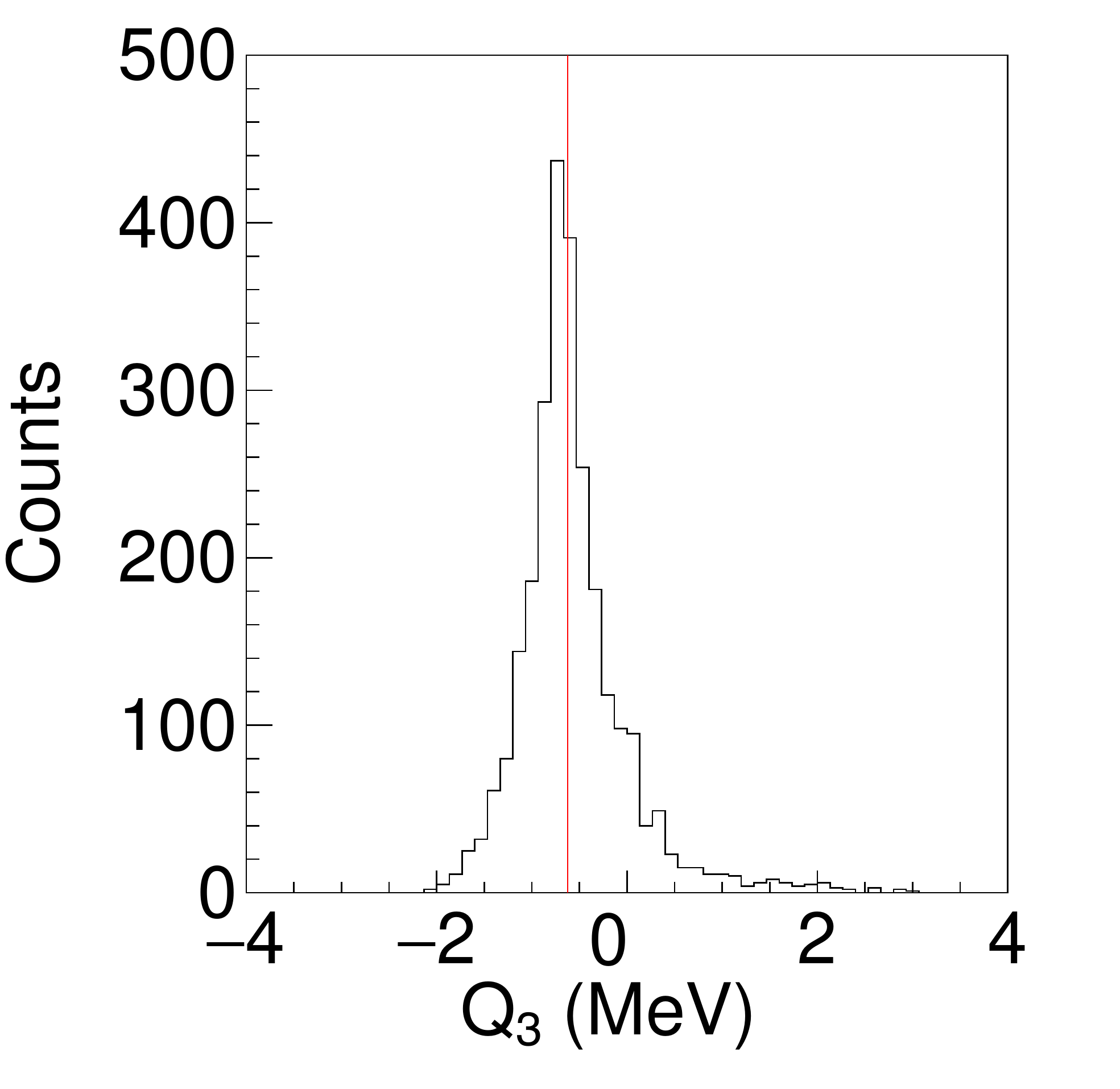}}
\caption{
THM reaction yield as a function of the calculated Q-value. The vertical red line points to the theoretical Q-value  of the three-body $^2$H$(^{27}$Al,$\alpha^{24}$Mg)n  reaction  (Q$_3$ = -0.6238 MeV).}
\label{qvalue}       
\end{figure}
This is why we did not use $\Delta$E to select $\alpha$-particles; however,
this makes it necessary to investigate reaction kinematics to see if 
the implementation of coincident detection in PSD1-PSD4 and PSD2-PSD3,
besides the selection of Z=12 nuclei in the telescopes, is
enough to suppress the contribution of background channels. 
To this purpose, for each angular couple $\theta_{{\rm PSD1}}-\theta_{{\rm PSD4}}$ and $\theta_{{\rm PSD2}}-\theta_{{\rm PSD3}}$ we have compared the
experimental data with a Monte Carlo simulation accounting for reaction kinematics only.
Indeed, events from the $^2$H$(^{27}$Al,$\alpha^{24}$Mg)n three-body reaction
should gather along characteristic loci in the ${\rm E}_{{\rm PSD1}}-{\rm E}_{{\rm PSD4}}$ and ${\rm E}_{{\rm PSD2}}-{\rm E}_{{\rm PSD3}}$ spectra. Fig.\ref{kinloc} shows such comparison for the angular condition $4.1{^\circ}\le\theta_{{\rm PSD1}}\le5.1{^\circ}$ and $27{^\circ}\le\theta_{{\rm PSD4}}\le30{^\circ}$, where the experimental THM data  (in black), obtained just by gating on the Z=12 locus in the $\Delta$E-E spectrum, are juxtaposed to the simulated spectrum, obtained through the Monte Carlo code mentioned above. 
Similar results, in good agreement with the simulation, are obtained also for other angular couples and for the PSD2-PSD3 detectors couple. The inspection of the spectra  makes it apparent that no background processes are populating the experimental spectra.
This is also apparent from the experimental Q-value spectrum
deduced from the THM reaction yield. It is shown in Fig.\ref{qvalue}
as an histogram, while the calculated Q-value of the $^2$H$(^{27}$Al,$\alpha^{24}$Mg)n reaction is shown as a vertical red line. Clearly, a single peak is present, excluding the occurrence of processes other than the $^2$H$(^{27}$Al,$\alpha^{24}$Mg)n reaction of interest. The agreement with the theoretical Q-value is also a positive test of the accuracy of the energy and angular calibrations we performed.

\section{Investigation of the QF mechanism}

\begin{figure}
\resizebox{0.48\textwidth}{!}{
\includegraphics{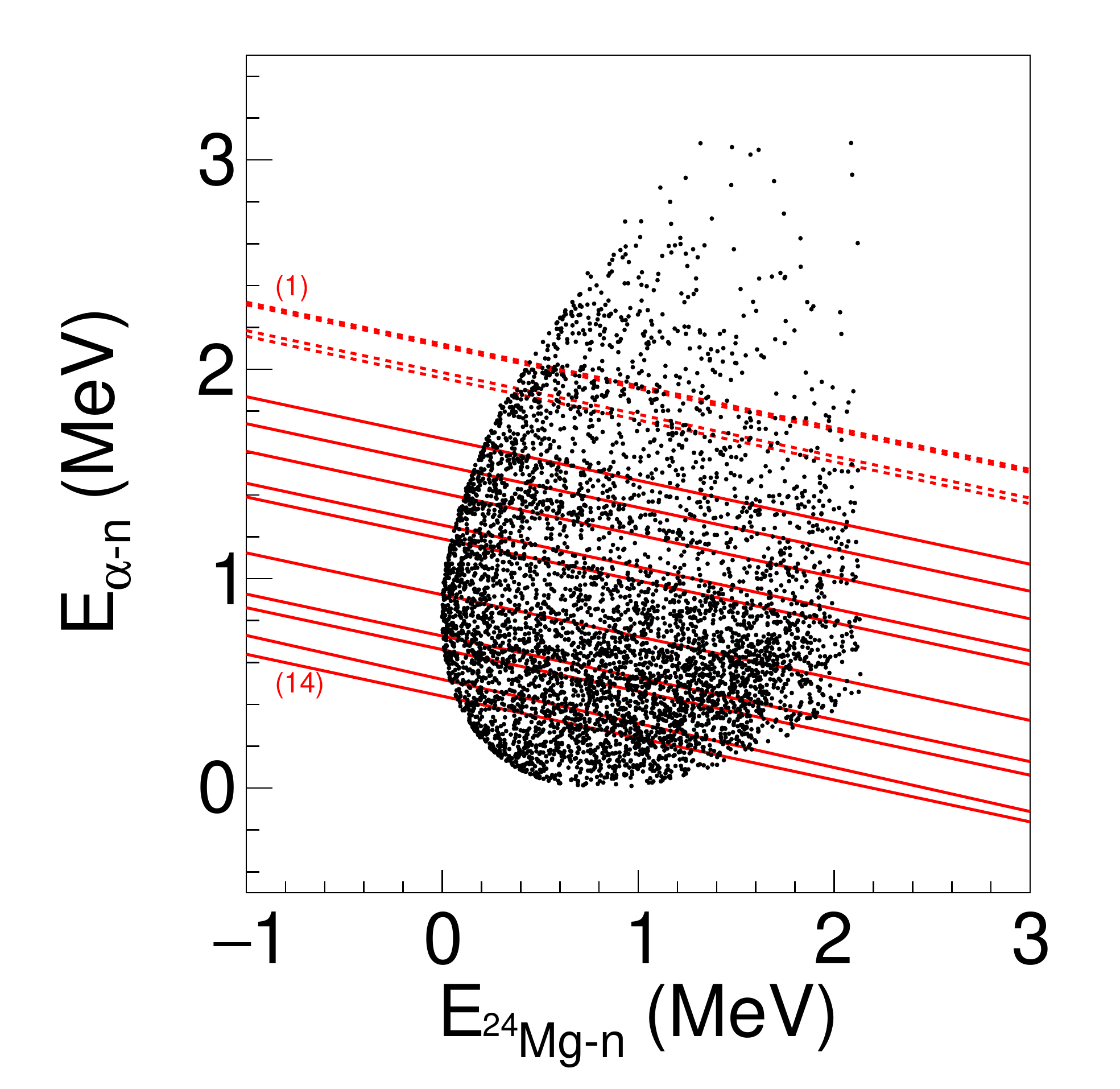}}
\caption{$^4$He-n vs. $^{24}$Mg-n relative energy 2D spectrum.
It is obtained imposing only the Z=12 condition on the 
$\Delta$E-E spectrum. The red lines are the $^{28}$Si levels listed
in \cite{ili10} with highest strength. Dashed lines
are used for those states for which only upper limits to
the resonant strengths could be set. The list of levels
marked in the picture is given in Tab.\ref{tablev}.}      
\label{erel}       
\end{figure}

The THM formalism for the extraction of the resonance strengths
requires the occurrence of the QF mechanism has been established, 
and the QF reaction yield separated from other reaction 
mechanisms. A first test is the study of the relative energy spectra
to rule out the occurrence of the sequential decay processes.
Indeed, the $^2$H$(^{27}$Al,$\alpha^{24}$Mg)n three-body reaction
might be described as a two step process, where $^5$He or $^{25}$Mg
are formed, later emitting a neutron to populate the 
$\alpha + {}^{24}$Mg+n exit channel (the case of the formation
of $^{28}$Si as intermediate step will be considered later on).
Indeed, if the neutron is emitted following the formation of an
intermediate system, deuteron breakup cannot be direct
and the $^2$H$(^{27}$Al,$\alpha^{24}$Mg)n is not proceeding
through a QF process. 

\begin{table}
\caption{List of the levels showing the highest strength in the
$^{27}$Al(p,$\alpha$)$^{24}$Mg reaction below 1.6~MeV.
States for which only an upper limit for the strengths is
available are reported in italic.  Resonance parameters (resonance 
energies, strengths and their uncertainties) are
taken from \cite{ili10}.}
\label{tablev}       
\centering
\begin{tabular}{llll}
\hline\noalign{\smallskip}
& E$_{{\rm c.m.}}$ (keV) & $\omega\gamma$ (eV) & $\delta\omega\gamma$ (eV) \\
\noalign{\smallskip}\hline\noalign{\smallskip}
(1) & 71.5   & $\le 2.47 \times 10^{-14}$ & - \\
(2) & 84.3   & $\le 22.60 \times 10^{-13}$ & - \\
(3) & 193.5  & $\le 23.74 \times 10^{-7}$ & - \\
(4) & 214.7  & $\le 21.13 \times 10^{-7}$ & - \\
(5) & 486.74 & 0.11                  & 0.05 \\
(6) & 609.49 & 0.275                 & 0.069 \\
(7) & 705.08 & 0.52                  & 0.13 \\
(8) & 855.85 & 0.83                  & 0.21 \\
(9) & 903.54 & 4.3                   & 0.4 \\
(10) & 1140.88 & 79                   & 27 \\
(11) & 1316.7 & 137                   & 47 \\
(12) & 1388.8 & 54                    & 15 \\
(13) & 1519.4 & 12.5 & 2.5 \\
(14) & 1587.87 & 30.0 & 6.0 \\
\noalign{\smallskip}\hline
\end{tabular}
\end{table}

To test this hypothesis, we have extracted 
the $^4$He-n vs. $^{24}$Mg-n relative energy spectrum, which
is shown in Fig.\ref{erel}. The occurrence of 
horizontal or vertical loci in the spectrum would signal the
formation of $^5$He or $^{25}$Mg intermediate nuclei, respectively, 
since the sum of the $^4$He-n vs. $^{24}$Mg-n relative energy
with the corresponding neutron emission threshold equals
the $^5$He or $^{25}$Mg excitation energies. Now, Fig.\ref{erel}
clearly rules out the formation of such states 
above about E$_{\alpha-n} \sim 1$~MeV,
no horizontal or vertical loci being observed in the experimental
spectra. Conversely, sloping loci are observable and
energy conservation consideration suggests to attribute them to 
the formation of $^{28}$Si excited states. In particular,
red lines are used to highlight the  $^{28}$Si states that are reported 
to have the highest strength for $^{27}$Al-p relative energies
less than about 1.5~MeV (dashed lines are used for states for
which upper limits are available only). The list of levels
marked in the figure is given in Tab.\ref{tablev}.
Below about E$_{\alpha-n} \sim 1$~MeV, a cluster of events
is present, which is not clearly attributable to 
sequential processes due to the formation of $^5$He
or $^{25}$Mg compound systems. Therefore, background
might be expected at high $\alpha-{}^{24}$Mg relative energies. 
\begin{figure}
\resizebox{0.48\textwidth}{!}{
\rotatebox{90}{\includegraphics{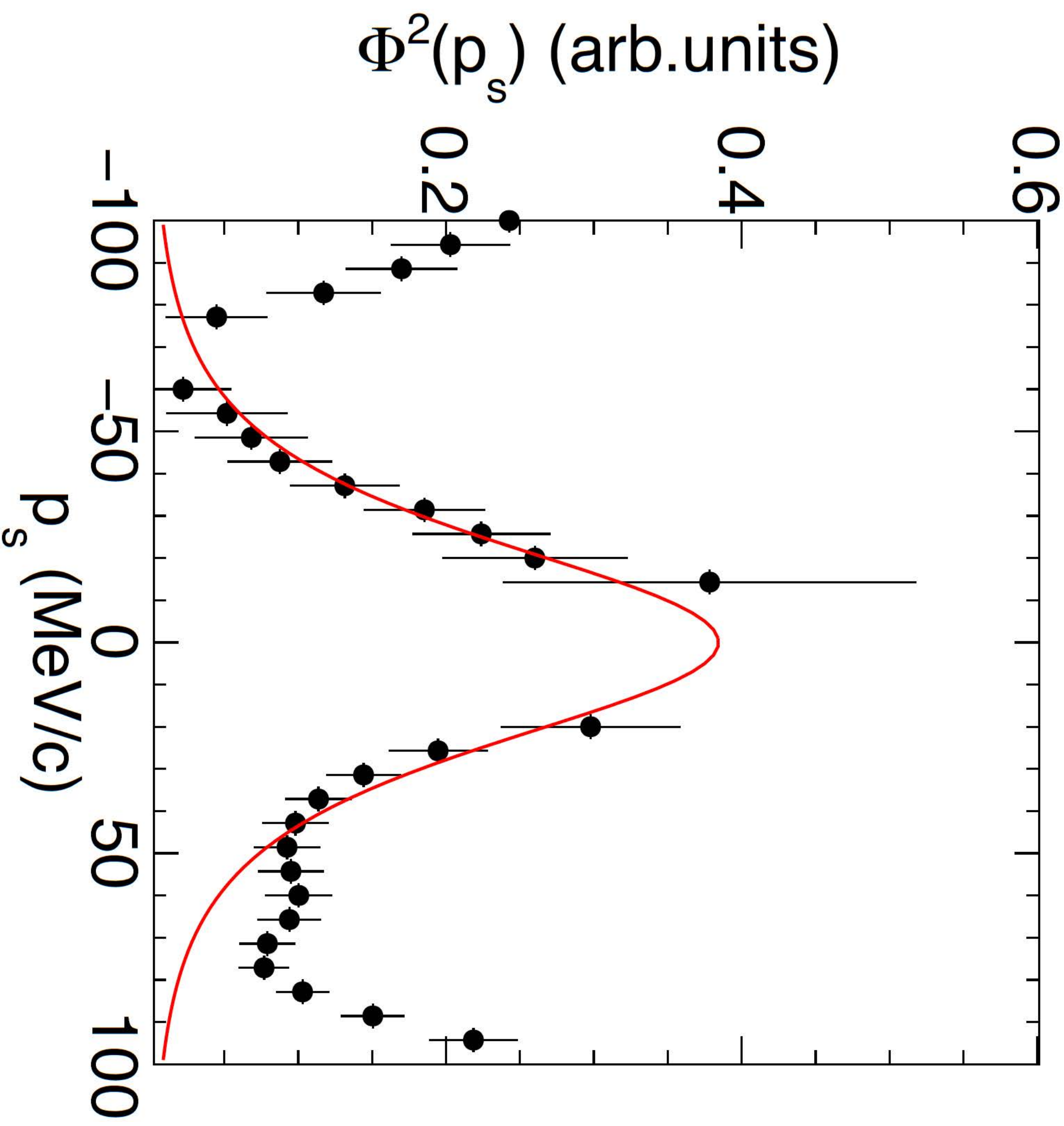}}}
\caption{Solid symbols: experimental neutron momentum distribution obtained introducing, besides the data analysis cut, a $1.1-1.2$~MeV gate on the $^{27}$Al-p relative energy. The red line is the neutron momentum distribution in deuteron as in \cite{lamia12}, normalized to the experimental data.}      
\label{ps2}
\end{figure}
Since  $^{28}$Si states may be populated by direct (QF) or
sequential process, further tests are necessary to establish
if the $^2$H$(^{27}$Al,$\alpha^{24}$Mg)n reaction proceeds
through QF mechanism. A procedure that has turned out to be
very effective is the extraction of the experimental neutron 
momentum distribution. Indeed, if the reaction mechanism is
direct, the neutron momentum distribution should be the same
as inside deuteron, the breakup process being adiabatic. Therefore,
if the experimental momentum distribution follows the theoretical
one, given by the squared deuteron wave function in momentum
space \cite{lamia12}, a necessary condition for the occurrence of 
the QF mechanism would be satisfied. 
From an experimental point of view, the neutron momentum 
distribution is obtained from the THM reaction yield by 
inverting Eq. \ref{thmeq}, namely, dividing it by
the kinematic factor under the assumption of a constant
HOES cross section. This is obtained by selecting 
a narrow energy (and angular) cut, such that in this
interval the variation is negligible. In the present case,
this was accomplished by introducing a 100~keV wide energy cut
around 2.75~MeV  $^{24}$Mg$-\alpha$ relative energy.
The resulting experimental neutron momentum distribution
is shown in Fig.\ref{ps2} as black solid circles,
while the theoretical trend given by the squared
Hulth\'en function in momentum space, as discussed in
\cite{lamia12}, is represented by a red line. The theoretical line was scaled 
to match the normalization of the experimental data.
From this figure, it is apparent that below about
${\left| \rm p_s \right| } \sim 50$~MeV/c, the agreement between the experimental and theoretical trend is very good, as it should be expected
(see \cite{spitaleri2019} for an extensive discussion). 
Similar results are obtained for other energy cuts.
This makes us confident about the occurrence of
the QF mechanism in the measured reaction and suggests the
most suitable cut to introduce in the next steps to single out
the QF yield. 

\section{Extraction of the E$_{c.m.}$ spectrum and final remarks}

\begin{figure}
\resizebox{0.48\textwidth}{!}{
\includegraphics{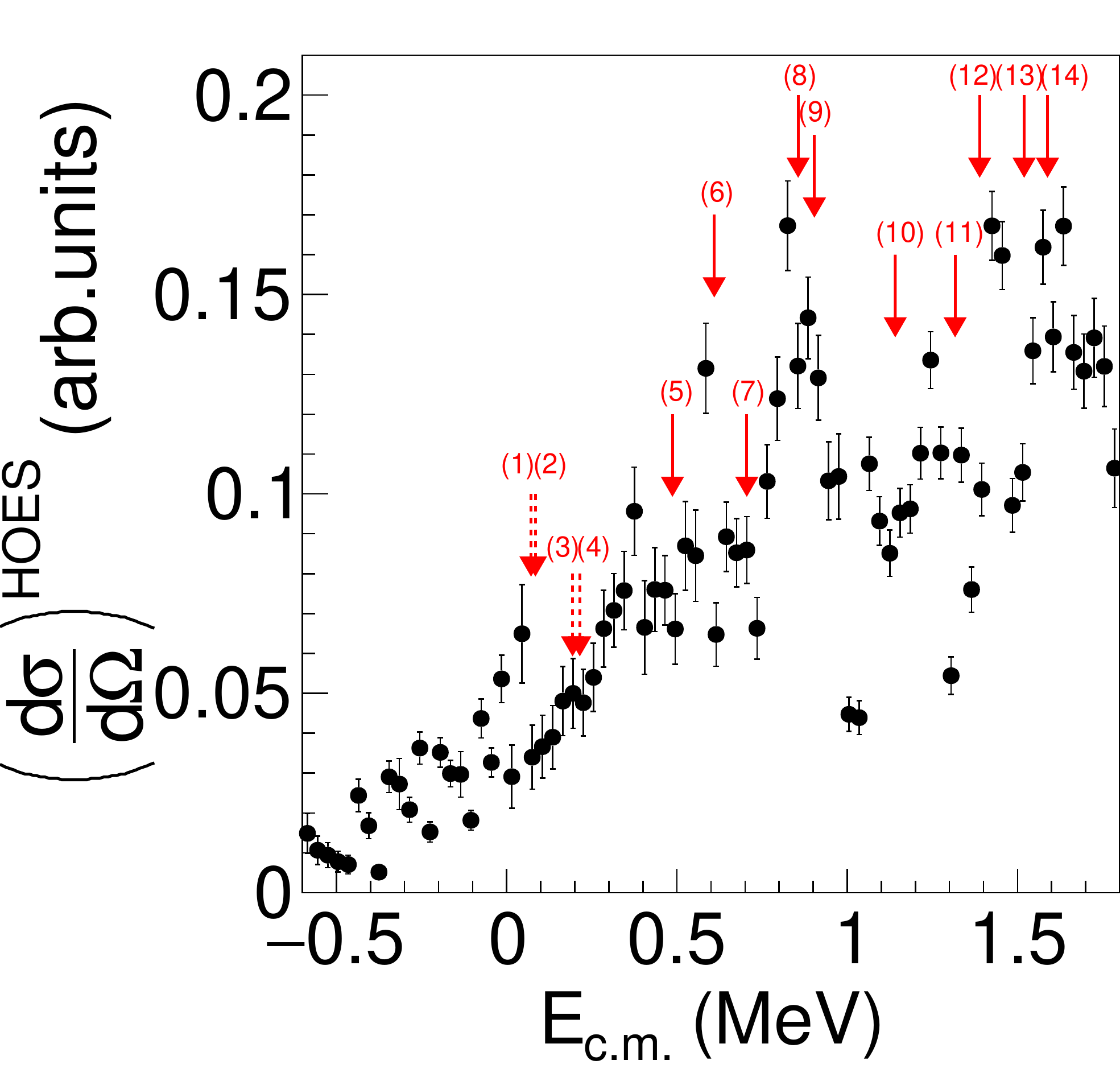}}
\caption{HOES 
$^{27}$Al(p,$\alpha$)$^{24}$Mg cross section $\left(\frac{d\sigma}{d\Omega}\right)^{HOES}$ 
as a function
of E$_{c.m.}$. Arrows mark the states listed in Tab.\ref{tablev}.
Dashed lines are used in the case only upper limits are available, as in Fig.\ref{erel}.
The spectrum was deduced taking $|p_s| \leq 50$~MeV/c.}       
\label{ecm}       
\end{figure}

From the previous analysis, we will consider only events
for which $|p_s| \leq 50$~MeV/c, the QF condition being 
satisfied and then Eq.\ref{thmeq} being applicable.
Fig.\ref{ecm} shows the QF coincidence yield, projected
on the ${\rm E_{c.m.}}$ variable, calculated as follows:
\begin{equation}\label{ecmeq}
{\rm E_{c.m.}=E_{^{24}Mg-\alpha}-Q_2}    
\end{equation}
where ${\rm E_{^{24}Mg-\alpha}}$ is the $^{24}$Mg$-\alpha$
relative energy and { \rm $Q_2$} is the Q-value of the 
$^{27}$Al(p,$\alpha$)$^{24}$Mg reaction \cite{spitaleri2019}. 
The coincidence yield is divided by the product 
$(KF)\mid\Phi(p_s)\mid^2$ to obtain the HOES 
$^{27}$Al(p,$\alpha$)$^{24}$Mg
cross section $\left(\frac{d\sigma}{d\Omega}\right)^{HOES}$, that is, 
by inverting Eq.\ref{thmeq}.

The obtained spectrum demonstrates the occurrence of a rich
pattern of resonances (highlighted by the red arrows) among which four resonant states sit right at astrophysical energies (below $\sim 0.2$~MeV). These latter are indicated by dashed lines because only upper limits are available for them in literature.
{ In detail, several resonance groups are visible; starting from the 
high energy edge there is a peak linked to the population of resonances (12-14),
with the state (12) probably well resolved from (13-14).
Then, a peak due to states (10-11) and another strong peak for states
(8-9) is visible, still populating energies above those of astrophysical
interest. At lower energies, there is an excess of events with respect to a 
smooth background around $\sim 0.5$~MeV, probably linked to the population
of states (5-7). Such bump is clearly seen even changing the energy bins.
Focusing on the low energy region, the one mostly affecting astrophysical 
consideration, the situation is more ambiguous since an increase in the energy
bin size would make the candidate peak at $\sim 0$~MeV disappear, so we cannot
draw astrophysics considerations at present.}

Indeed, if  these resonance strengths (and of the one at 71.5 or at 84 keV in particular) were found greater than the established upper limit, the $^{27}$Al(p,$\alpha$)$^{24}$Mg reaction rate would be larger than the $(p,\gamma)$ competitor channel at astrophysical energy. The Gamow window of the $^{27}$Al(p,$\alpha$)$^{24}$Mg reaction at the average temperature of stellar H-burning (T$_9\sim$ 0.055) ranges between 70 and 120 keV and the narrow temperature range between T$_9=$0.07 and 0.08, where the fate of $^{24}$Mg nucleosynthesis in stars is decided, spans from 80 to 160 keV. 
Moreover, if the rate of the $^{27}$Al(p,$\alpha$)$^{24}$Mg would be larger, the MgAl cycle will turn to be well closed with a larger production of $^{24}$Mg and a more efficient destruction of $^{27}$Al, even in low mass stars, and then the problem of the overproduction of $^{26}$Al with respect to $^{27}$Al in H-burning environments could be  alleviated.

Unfortunately, owing to low statistics, it is not possible
to extract angular distributions for each state. Indeed, 
binning of the coincidence yield limits the resolution, 
so several resonances often overlap. Therefore, we could not
carry out angular integration, as discussed in  \cite{indelicato2017}.
Such integration is a pivotal step in the application of
the procedure sketched in \cite{laco10}, to extract the resonance
strengths from the peak areas. An especially important
point to be addressed is the increase of $\left(\frac{d\sigma}{d\Omega}\right)^{HOES}$ with increasing E$_{c.m.}$,
which is probably due to the unresolved contribution of the
sequential decays pointed out when discussing Fig.\ref{erel}.

In any case, this work has made it possible
to establish the occurrence of the QF mechanism in the 
$^2$H$(^{27}$Al,$\alpha^{24}$Mg)n reaction, which is a preliminary
test for the application of the method. 
A future dedicated experiment with 
higher statistics would make it possible to explore the energy
region below about 500~keV and determine the strengths of the observed 
levels at about 80~keV and at about 200~keV (see Tab.\ref{tablev}
for details). Also, the possibility to cover the energy region
above about 500~keV where additional resonances are apparent, 
in particular around 900~keV (see Fig.\ref{erel}),
would make it possible to perform normalization of the 
resonant strengths as discussed in \cite{laco15}. 

\section{Acknowledgements}
The authors thanks the LNS technical staff for the support during the experimental run. 
V. Burjan,  J. Mrazek and G. D’Agata acknowledge the support from MEYS Czech Republic under the project EF16 013/0001679. G. G. Kiss acknowledges support from the NKFIH (NN128072), the UNKP-20-5-DE-2 New National Excellence Program of the Ministry of Human Capacities of Hungary, and the Janos Bolyai research fellowship of the Hungarian Academy of Sciences. L. Acosta and E. Ch\'avez  acknowledge the support from DGAPA-UNAM IN107820, AG101120 and CONACyT 314857.
%
%

\end{document}